\documentclass[aps,prl,showpacs,twocolumn,groupedaddress]{revtex4}

\usepackage{graphics}
\usepackage{epsfig}
\usepackage{mathrsfs}
\usepackage{amsbsy}
\usepackage{amssymb}

\newcommand{\half}{\mbox{$\frac12$}}

\begin{document}

\title{Transient Growth in Shear Flows:\\Linearity vs Nonlinearity}

\author{Chris C.T. Pringle}
\email{C.C.T.Pringle@reading.ac.uk}
\author{Rich R. Kerswell}
\email{R.R.Kerswell@bristol.ac.uk}

\affiliation{Department of Mathematics, University of Bristol, University Walk,
   Bristol BS8 1TW, United Kingdom}
\date{\today}

\begin{abstract}
Two approaches to the problem of transition to turbulence of shear
flows are popular in the literature. The first is the linear one of
transient growth which focuses on the likely {\em form} of the most
`dangerous' (lowest energy) turbulence-triggering disturbances. The
second is the nonlinear calculation of the laminar-turbulent
boundary which instead focuses on their typical {\em amplitudes}. We
look to bridge the gap between these two perspectives by considering
the fully nonlinear transient growth problem to estimate both the
form and amplitude of the most dangerous disturbance. We thereby
discover a new nonlinear optimal disturbance which outgrows the
well-known linear optimal for the same initial energy and is
crucially much more efficient in triggering turbulence. The
conclusion is then that the most dangerous disturbance can differ
markedly from what traditional linear transient growth analysis
predicts.
\end{abstract}

\maketitle

Shear flows are ubiquitous in nature and engineering, and
understanding how and why they become turbulent has huge economic
implications.  This has led to a number of simplified canonical
problems being studied such as plane Couette flow, channel flow and
pipe flow which commonly exhibit turbulent behavior even when the
underlying laminar state is linearly stable.  In this case, a finite
amplitude perturbation is required in order to trigger turbulence and
a leading question is then what is the `most dangerous' or `smallest' 
such perturbation (with the metric typically being energy). Beyond its
intrinsic interest, such information is fundamentally important for
devising effective control strategies to delay the onset of turbulence.

Historically, linear transient growth analysis
\cite{tref,schmid,gustav,butler,berg} has been used to identify
dangerous disturbances which are efficient at triggering
turbulence. This focuses on a linear mechanism whereby infinitesimally
small perturbations can interact with the underlying shear profile in
order to create much more energetic disturbances. Ultimately, these
disturbances ebb away if the shear profile is linearly stable but the
thinking is that the large growth possible can catapult the
disturbance into a regime where nonlinear effects sustain its energy
away from zero for all times.

Alternatively, recent progress has been made \cite{itano,schneider} in
numerically tracking the laminar-turbulent boundary which represents
the surface in phase space which separates those initial conditions
which will trigger a turbulent episode from those which will simply
relaminarise. Since the tracking technique hinges upon carefully
selecting initial conditions and integrating forward in time, the part
of the boundary revealed is effectively confined to the neighborhood
of the limiting set of the boundary-confined flow dynamics. This is
found to have significantly higher energy levels than that actually
needed to trigger turbulence by carefully-tuned initial disturbances
(e.g. \cite{viswanath}). This disparity in energy levels is none other
than an expression of the large transient growth endemic in shear
flows but now crucially translated into the nonlinear regime.  The
most dangerous disturbance corresponds with the minimum energy point
on the laminar-turbulent boundary and stands to gain the largest
energy as it sweeps up to the limiting set energy plateau. In this
Letter, we pursue the promising strategy of extending the usual linear
transient growth analysis into the nonlinear regime to identify the
most dangerous disturbance in pipe flow. Our findings should be
equally relevant to other shearing flows such as plane Couette flow,
channel flow, boundary layers, etc.

The transient growth problem is the optimisation question: what
initial condition $\mathbf{u}(\mathbf{x},t=0)$ (added as a
perturbation to the laminar flow $2(1-4s^2)\hat{\mathbf{z}}\,$) for the
governing Navier-Stokes equations with fixed (perturbation) kinetic
energy $E_0$ will give rise to the largest subsequent energy $E_T$ at
a time $t=T$ later. This corresponds to maximising the functional
\begin{eqnarray}
\mathscr{L} &:=& \langle \half
\mathbf{u}(\mathbf{x},T)^2\rangle-\lambda\langle \half
\mathbf{u}(\mathbf{x},0)^2-E_0\rangle \nonumber \\ 
&&-\int_0^T \langle \boldsymbol{\nu},\bigg[\frac{\partial \mathbf{u}}{\partial t} 
-16su\hat{\mathbf{z}}+2(1-4s^2)\frac{\partial \mathbf{u}}{\partial z} \nonumber\\
&& \qquad \qquad \qquad +\mathbf{u}\cdot\nabla\mathbf{u}
+ \nabla p - \frac{1}{Re} \nabla^2 \mathbf{u} \bigg] \rangle dt
\nonumber \\
&&-\int_0^T \langle \Pi \nabla \cdot \mathbf{u} \rangle dt 
-\int_0^T \Gamma \langle \mathbf{u} \cdot \mathbf{\hat{z}} \rangle dt
\end{eqnarray}
where $\langle \, \, \rangle$ represents volume integration;
$(s,\phi,z)$ are cylindrical coordinates directed along the pipe;
$\lambda$, $\boldsymbol{\nu}(\mathbf{x},t)$, $\Pi(\mathbf{x},t)$ and
$\Gamma(t)$ are Lagrange multipliers imposing the constraints of
initial energy $E_0$, that the Navier-Stokes equations hold over $t
\in [0,T]$, incompressibility and constant mass flux in time
respectively (the system has been non-dimensionalised by the pipe
diameter $D$ and the bulk velocity $U$ with $Re:=\rho UD/\mu$ where
$\rho$ is the density and $\mu$ is the dynamic viscosity). Vanishing
of the variational derivatives requires that $\mathbf{u}$ must evolve
according to the Navier-Stokes equations, $\boldsymbol{\nu}$ evolves
according to the adjoint-Navier-Stokes equations and at times $t=0$
and $T$ we have optimality and compatibility conditions linking the
two sets of variables (e.g. see \cite{guegan} for details of the
linearised problem). The method of solution is one of iteration as
follows:
\begin{itemize}
\item Make an initial guess for $\mathbf{u}(\mathbf{x},t=0)$.
\item Allow $\mathbf{u}(\mathbf{x},t)$ to evolve according to the
  Navier-Stokes equations until $t=T$.
\item Solve the compatibility condition for
  $\boldsymbol{\nu}(\mathbf{x},T)$, $ \delta \mathscr{L}/\delta
  \mathbf{u}(\mathbf{x},T) \equiv
  \mathbf{u}(\mathbf{x},T)-\boldsymbol{\nu}(\mathbf{x},T) =
  \mathbf{0}$
\item Allow the incompressible field $\boldsymbol{\nu}(\mathbf{x},t)$
  to evolve \emph{backwards} in time until $t=0$ via the
  adjoint-Navier-Stokes equations
\begin{eqnarray}
&& \frac{\partial \boldsymbol{\nu}}{\partial t} + 2(1-4s^2)\frac{\partial \boldsymbol{\nu}}{\partial s} 
+ \frac{1}{s}(\nu_\phi u_s - \nu_s u_\phi)\boldsymbol{\hat{\phi}}  
 + \mathbf{u} \cdot \nabla) \boldsymbol{\nu} \nonumber \\ 
&&\qquad + 16s\nu_3\mathbf{\hat{s}}+(u_i \partial_j \nu_i)
= -\nabla \Pi - \frac{1}{Re} \nabla^2 \boldsymbol{\nu}
\end{eqnarray} 
\item Move $\mathbf{u}(\mathbf{x},0)$ in the direction of the variational
  derivative $ \delta \mathscr{L}/\delta
  \mathbf{u}(\mathbf{x},0) \equiv
  -\lambda\mathbf{u}(\mathbf{x},0)+\boldsymbol{\nu}(\mathbf{x},0)$ and repeat.
\end{itemize}

Both direct and adjoint equations were solved using a fully spectral,
primitive variables approach.  Time stepping was done using a second
order fractional step scheme, checked carefully against the code of
\cite{willis}. The computational domain was a short periodic domain of
length $\pi$ radii with typical spatial resolution of 29 real Fourier modes
azimuthally, 11 real Fourier modes axially and 25 modified Chebyshev
polynomials radially in each of the 8 physical scalar fields
$(u,v,w,p,\nu_1,\nu_2,\nu_3,\Pi)$. 
All results have been checked for robustness to resolution changes.
Retention of the nonlinear terms poses a fresh technical challenge: although 
the adjoint equation is linear in $\mathbf{\nu}$, it is dependent on 
the evolution history of the forward variable $\mathbf{u}$ which now must 
be stored.

\begin{figure}
 \begin{center}
  \includegraphics[width=\columnwidth]{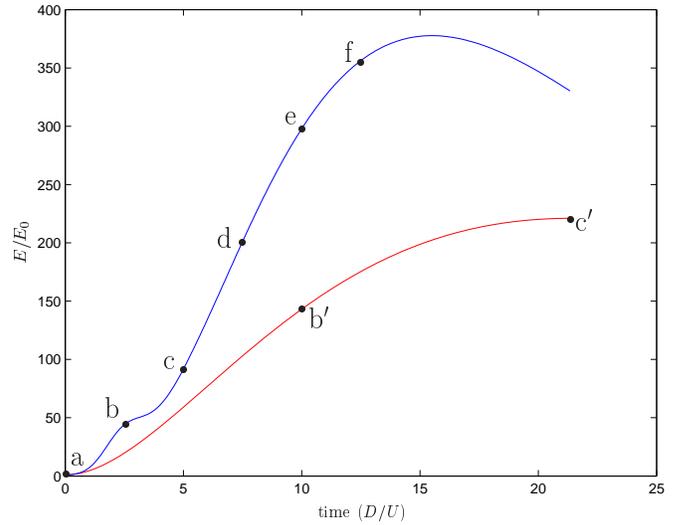}
  \caption{The evolution of the linear and nonlinear optimals at
    $Re=1750$. The blue line corresponds to the nonlinear optimal for
    $E_0=2\times 10^{-5}$ while the red line is the linear optimal
    ($E_0 \rightarrow 0$). The nonlinear result produces more growth
    and actually reaches its maximum at a slightly earlier time than
    $T$. \label{nl_opt}}
 \end{center}
\end{figure}

\begin{figure}
 \begin{center}
  \includegraphics[width=\columnwidth]{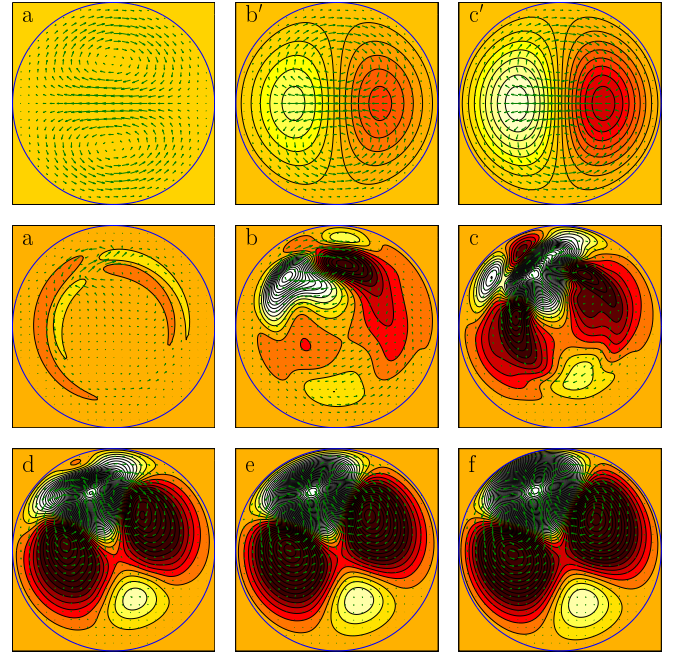}
  \caption{Three snapshots of the linear optimal (top) and six
    snapshots (middle \& bottom) of the 3D optimal for $Re=1750$ and
    $E_0=2\times 10^{-5}$ during its evolution.  Labels refer to
    figure \ref{nl_opt}, arrows indicate cross-sectional velocities
    and colours axial velocity beyond the laminar flow (white/light
    for positive and red/dark for negative: outside shade represents
    zero).\label{nl_evol}}
 \end{center}
\end{figure}
\begin{figure}

 \begin{center}
  \includegraphics[width=\columnwidth]{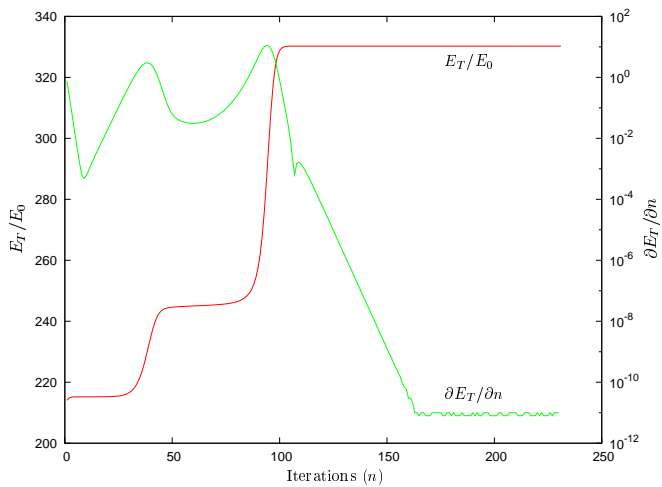}
  \caption[Convergence of the nonlinear optimal]{$Re=1750, E_0=2\times
    10^{-5}$. The iterations are seeded with a noisy version of the 2D
    optimal and, after approaching two `saddle' points, eventually
    converge onto the 3D optimal.\label{fig:saddle3_conv}}
 \end{center}
\end{figure}
%
% Results
%

The linear transient growth optimal $\mathbf{u}_{lin}(\mathbf{x};Re)$
in pipe flow is well-known to be streamwise-independent (2D) rolls
which evolve into much larger streamwise-independent streaks
\cite{schmid}: see figs 1 and 2. Maximum growth occurs at
$T_{lin}\approx 12.2 \times Re/1000\,(D/u)$ \cite{mes}.  Introducing
nonlinearity (ie increasing $E_0$ from $0$), setting $T=12.2 \times
Re/1000\,(D/U)$ and allowing only 2D flows, leads smoothly to a
modified 2D optimal $\mathbf{u}_{2D}(\mathbf{x};E_0,Re)$ with
monotonically decreasing growth (see \cite{zuccher} for an equivalent
result in boundary layers). By $E_0 \sim 10^{-3}$ (in units of the
laminar flow's kinetic energy, used henceforth), the energy
magnification falls to about $50\%$ of its linear value at $Re=1750$,
and continues to decrease thereafter. Opening the optimisation up to
fully 3D flows initially just recovers the 2D result but once $E_0$
crosses a small threshold $E_{3D}$ ($\approx 10^{-5}$ at $Re=1750$), a
completely new optimal $\mathbf{u}_{3D}(\mathbf{x};E_0,Re)$ appears.
This new 3D optimal emerges from the optimisation procedure after it
initially appears to converge to the 2D optimal and then transiently
visits an intermediate state: see fig 3. Identifying this `loss of
stability' of the 2D optimal provided an efficient way to compute
$E_{3D}(Re)$.  All optimisation results were robust over three very
different choices of starting flow: a) $\mathbf{u}_{lin}$ with noise;
b) the asymmetric travelling wave \cite{PK07}; and c) a turbulent flow
snapshot (all rescaled to the appropriate initial energy).

Given the intensity of the runs ($O(200)$ iterations and each
iteration requires integrating forwards and backwards over the period
$[0,T]$), we concentrated on two values of the Reynolds number,
$Re=1750$ and $2250$, and the corresponding energy ranges
$[E_{3D}=1.35\times 10^{-5},2\times 10^{-5}]$ and $[E_{3D}=4.8\times
  10^{-6}, 6.25\times 10^{-6}]$. Over both intervals the 3D optimal
has essentially the same appearance: see fig. \ref{nl_evol}.  Unlike
the linear optimal which is globally simple in form and undergoes an
evolution that is well established (rolls advecting the mean shear to
generate streaks), the 3D optimal is localised to one side of the pipe
and initially has both rolls and streaks of comparable
amplitude. Figures \ref{nl_opt} and \ref{nl_evol} show a new 2-stage
evolution: a preliminary phase when the flow delocalises followed by a
longer growth phase where the flow structure stabilises to essentially
two large-scale slow streaks sandwiching one fast streak near the
boundary.

\begin{figure}
 \begin{center}
  \includegraphics[width=\columnwidth]{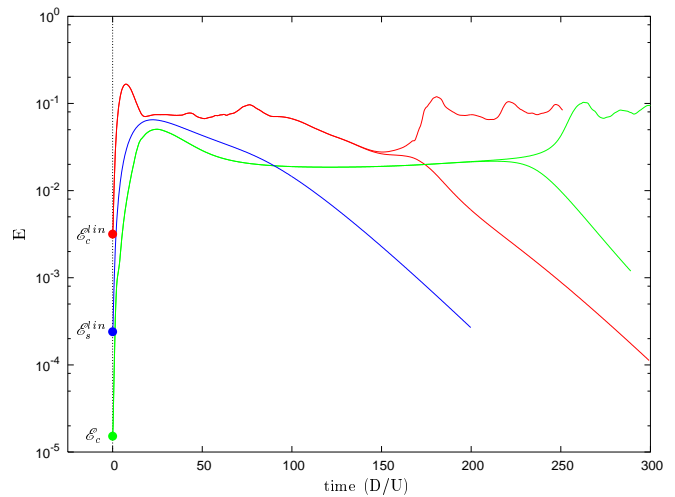}
  \caption{Re=2500. The green line shows the evolution of the 3D
    optimal when given initial energy $\mathscr{E}_c$. Because it is
    on the laminar-turbulent boundary two nearly identical initial
    conditions diverge after, in this case, $220D/U$. The blue line is
    the evolution of the 2D optimal for the exact initial energy
    $\mathscr{E}_s^{lin}$ for which the streaks become linearly
    unstable.  The red line shows the 2D optimal given initial energy
    $\mathscr{E}_c^{lin}$ and allowed to evolve until it reaches a
    maximum amplitude whereupon $0.1\%$ by amplitude unstable
    perturbation is added. Again the laminar-turbulent boundary can be
    identified.\label{edges}}
 \end{center}
\end{figure}

\begin{figure}
 \begin{center}
  \includegraphics[width=\columnwidth]{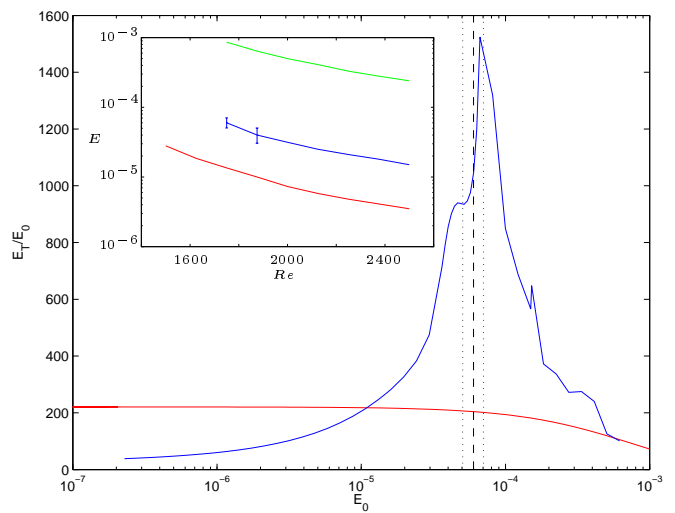}
  \caption{The effect of the initial energy on the growth of
    $A\mathbf{u}_{lin}$ (red) and $A\mathbf{u}_{3D}$ (blue) at
    $Re=1750$. For small $E_0$ the 2D result is the optimal but after
    $E_{3D}=1.35 \times 10^{-5}$, the 3D optimal takes over. The
    vertical dashed line corresponds to $\mathscr{E}_c$, with the
    dotted lines being the relevant errorbars.  \textbf{Inset:} The
    dependence of $E_{3D}$ (red), $\mathscr{E}_{c}$ (blue) and
    $\mathscr{E}_s^{lin}$ (green) on $Re$ ($\mathscr{E}_c^{lin}$ is
    even higher). The error bars shown on $\mathscr{E}_c$ at $Re=1750$
    and $1875$ are because the laminar-turbulent boundary is difficult
    to identify at these $Re$ due to the pipe shortness.  \label{amplitude}}
 \end{center}
\end{figure}

If $E_0$ is increased beyond the ranges quoted above, the iterative
procedure fails to converge for reasons which are unclear. One
possibility is that at these energy levels, the laminar-turbulent
boundary has been crossed and the lack of convergence is due to the
flow becoming turbulent. The ensuing sensitivity to noise would make
the optimisation non-smooth. However, direct numerical simulation
starting with the 3D optimal does not reveal a turbulent episode
implying that there is still an energy gap between where the 3D
optimal emerges as the solution to the transient growth problem and
the lowest energy of any initial condition which can trigger
turbulence \footnote{There is the assumption that the optimisation
  algorithm samples all possible flows of a given energy and thus if
  it converges smoothly, turbulence cannot be triggered at this energy
  level.}.  What can be shown, however, is that the 3D
optimal is much more efficient at triggering turbulence than the
linear optimal when rescaled. Taking the initial condition
$A\mathbf{u}_{3D}(\mathbf{x};2\times10^{-5},1750)$, we gradually
increase the rescaling factor $A$ until a critical energy
$E_0=\mathscr{E}_c(Re)$ is reached at which turbulence is triggered.
Calculating the corresponding quantity for the linear optimal turns
out to be less clearly defined because some 3D noise is needed to
trigger turbulence. As a result we make 2 different estimates, one
strictly conservative and the other more realistic. The first
$\mathscr{E}_s^{lin}$ is obtained by taking
$A\mathbf{u}_{lin}(\mathbf{x};1750)$ and finding the initial energy
for which the resultant streaks are just linearly unstable in this
periodic domain \cite{reddy}.  In the second $\mathscr{E}_c^{lin}$,
the same initial condition was used but $0.1\%$ of the most unstable
perturbation (as found from the previous computation) is added to the
streaks when they reach maximum amplitude. $\mathscr{E}_s^{lin}$
should be a (low) conservative estimate but even this is O(10) times
larger than $\mathscr{E}_c$ at $Re=2500$ - see figure \ref{edges} -
whereas the more realistic $\mathscr{E}_c^{lin}$ is O(100) times
larger.

In figure \ref{amplitude}(inset) we plot $E_{3D}$, $\mathscr{E}_c$ and
$\mathscr{E}_s^{lin}$ as a function of $Re$ which emphasizes that the
2D optimal (for which the linear result is an excellent approximation) 
ceases to be a global maximum at an energy (at least)
several orders of magnitude before it approaches the laminar-turbulent
boundary.  The 3D optimal, in contrast, crosses the laminar-turbulent
boundary only shortly after it emerges at $E_{3D}$. This indicates
that the 3D optimal provides a rapid means of bridging the gap between
when the 3D nonlinear optimal surpasses the linear result and when turbulence
can be triggered. To achieve this, the energy growth experienced by
the 3D optimal must increase dramatically with $E_0$ which is 
illustrated in figure \ref{amplitude}. It is worth remarking that the lowest
possible energy to trigger turbulence must be bounded below by
$E_{3D}$ [15] and above by
$\mathscr{E}_{c}$.
% CAREFUL - manual referencing to footnote!! - RRK 6-5-10

%
% Discussion
%

In this letter we have demonstrated that including nonlinearities in
the problem of transient growth critically changes the result close to
the onset of turbulence. Although we have not been able to calculate
these solutions all the way up to the laminar-turbulent boundary as
originally hoped, we provide evidence that they are very efficient at
triggering turbulent episodes, notably more so than the linear
result. Admittedly, we have only considered a short periodic domain
and so the natural question is what will happen in larger
domains. Here we expect further localisation of the optimal since
energy is defined as a {\em global} quantity whereas nonlinearity is
important whereever the velocity field is \emph{locally} large. This
strongly suggests that in a long pipe the optimal should localise in
the axial direction as well (the 3D optimal found here is already
localised in the radial and azimuthal directions). This squares well
with the experimental observation that small local perturbations can
trigger high energy global turbulence. More importantly, the fact that
localised flow structures should emerge from this type of theoretical
analysis bodes well for a greater connection between theory and
experiments which naturally introduce localised disturbances.

Finding fully nonlinear optimals is a time-consuming pursuit due to
the slow convergence of the iterative procedure and the need
to look within small energy windows. Their discovery has had to wait
almost two decades after the linear result was established in pipe
flow. As computer power steadily increases, we envision that these new
nonlinear optimals will start to come within easy reach. That the
optimal found here represents something entirely different to the
previously known linear results suggests that they will open up a
whole new means of triggering turbulence and a whole new way of
understanding how transition occurs.

\begin{acknowledgments}
  The calculations in this paper were carried out at the Advanced
  Computing Research Centre, University of Bristol.
\end{acknowledgments}

\end{document}